\documentclass[aps,prl,onecolumn,11pt,superscriptaddress]{revtex4-1}

\pdfoutput=1
 
\usepackage{graphicx}
\usepackage{amssymb,amsfonts,amsmath}
\usepackage{bm}
\usepackage{textcomp}
\usepackage[english]{babel}
\usepackage{url}
\usepackage{hyperref}

\def\b#1\e{\begin{align}#1\end{align}}	
\newcommand{\I}{i}

\newcommand{\ehoch}[1]{\operatorname{exp}{\left[ #1 \right]}}
\newcommand{\E}[1]{\operatorname{e}^{#1}}

\renewcommand{\vec}[1]{\ensuremath{\mathchoice{\mbox{\boldmath$\displaystyle#1$}}
                            {\mbox{\boldmath$\textstyle#1$}}
                            {\mbox{\boldmath$\scriptstyle#1$}}
                            {\mbox{\boldmath$\scriptscriptstyle#1$}}}}
\newcommand{\R}{\vec{\mathcal{R}}}
\newcommand{\sP}{\vec{\mathcal{P}}}

\begin{document}


\title{Interferometry with Bose-Einstein Condensates in Microgravity}



\def\affzarm {\affiliation{ZARM, Universit\"at Bremen, Am Fallturm, 28359 Bremen, Germany }}
\def\affiqo  {\affiliation{Institut f\"ur Quantenoptik, Leibniz Universit\"at Hannover, Welfengarten 1, 30167 Hannover, Germany }}
\def\affhub  {\affiliation{Institut f\"ur Physik, Humboldt-Universit\"at zu Berlin, Newtonstr. 15, 12489 Berlin, Germany }}
\def\affuhh  {\affiliation{Institut f\"ur Laser-Physik, Universit\"at Hamburg, Luruper Chaussee 149, 22761 Hamburg, Germany }}
\def\affulm  {\affiliation{Institut f\"ur Quantenphysik and Center for Integrated Quantum Science and Technology (IQST), Universit\"at Ulm, Albert-Einstein-Allee 11, 89081 }}
\def\affuk   {\affiliation{Midlands Ultracold Atom Research Centre, Birmingham B15 2TT, United Kingdom }}
\def\affDLR  {\affiliation{DLR Institut f\"ur Raumfahrtsysteme, Robert-Hooke-Str. 7, 28359 Bremen, Germany }}
\def\afftuda {\affiliation{Institut f\"ur Angewandte Physik, Technische Universit\"at Darmstadt, Hochschulstr. 4A, 64289 Darmstadt, Germany }}
\def\affMPQ  {\affiliation{Max-Planck-Institut f\"ur Quantenoptik und Fakult\"at f\"ur Physik der Ludwig-Maximilians-Universit\"at M\"unchen, Schellingstr. 4, 80799 M\"unchen, Germany }}
\def\affsaar {\affiliation{Theoretische Physik, Universit\"at des Saarlandes, Campus E2 6, D-66041 Saarbr\"ucken, Germany }}
\def\afffbh  {\affiliation{Ferdinand-Braun-Institut, Leibniz-Institut f\"ur H\"ochstfrequenztechnik, Gustav-Kirchhoff-Str. 4, 12489 Berlin, Germany }}
\def\affens  {\affiliation{Laboratoire Kastler Brossel, ENS/UPMC-Paris 6/CNRS, 24 rue Lhomond, 75005 Paris, France }}

\author{H.~M\"untinga}\thanks{The Authors contributed equally}
\affzarm
\author{H.~Ahlers}\thanks{The Authors contributed equally}
\affiqo
\author{M.~Krutzik}\thanks{The Authors contributed equally}
\affhub
\author{A.~Wenzlawski}\thanks{The Authors contributed equally}
\affuhh
\author{S.~Arnold}
\affulm 
\author{D.~Becker}
\affiqo
\author{K.~Bongs}
\affuk
\author{H.~Dittus}
\affDLR
\author{H.~Duncker}
\affuhh
\author{N.~Gaaloul}
\affiqo
\author{C.~Gherasim}
\afftuda 
\author{E.~Giese}
\affulm
\author{C.~Grzeschik}
\affhub
\author{T.~W.~H\"ansch}
\affMPQ
\author{O.~Hellmig}
\affuhh
\author{W.~Herr}
\affiqo
\author{S.~Herrmann}
\affzarm
\author{E.~Kajari}
\affulm
\affsaar
\author{S.~Kleinert}
\affulm
\author{C.~L\"ammerzahl}
\affzarm
\author{W.~Lewoczko-Adamczyk}
\affhub
\author{J.~Malcolm}
\affuk
\author{N.~Meyer}
\affuk
\author{R.~Nolte}
\afftuda
\author{A.~Peters}
\affhub
\afffbh 
\author{M.~Popp}
\affiqo
\author{J.~Reichel}
\affens
\author{A.~Roura}
\affulm
\author{J.~Rudolph}
\affiqo
\author{M.~Schiemangk}
\affhub
\afffbh
\author{M.~Schneider}
\afftuda
\author{S.~T.~Seidel}
\affiqo
\author{K.~Sengstock}
\affuhh
\author{V.~Tamma}
\affulm
\author{T.~Valenzuela}
\affuk
\author{A.~Vogel}
\affuhh
\author{R.~Walser}
\afftuda
\author{T.~Wendrich}
\affiqo
\author{P.~Windpassinger}
\affuhh
\author{W.~Zeller}
\affulm
\author{T.van Zoest}
\affDLR
\author{W.~Ertmer}
\affiqo
\author{W.~P.~Schleich}
\affulm
\author{E.~M.~Rasel}
\email[To whom correspondence should be addressed: ]{rasel@iqo.uni-hannover.de}
\affiqo


\date{\today}

\begin{abstract}
Atom interferometers covering macroscopic domains of space-time are a 
spectacular manifestation of the wave nature of matter. Due to their 
unique coherence properties, Bose-Einstein condensates are ideal sources 
for an atom interferometer in extended free fall. In this paper we report on 
the realization of an asymmetric Mach-Zehnder interferometer operated with a 
Bose-Einstein condensate in microgravity. The resulting interference pattern 
is similar to the one in the far-field of a double-slit and shows a linear 
scaling with the time the wave packets expand. We employ delta-kick cooling 
in order to enhance the signal and extend our atom interferometer. 
Our experiments demonstrate the high potential of interferometers operated 
with quantum gases for probing the fundamental concepts of quantum mechanics 
and general relativity. 
\end{abstract}


\maketitle


Quantum theory \cite{bohm_quantum_1951} and general relativity 
\cite{misner_gravitation_1973} are two pillars of
modern physics and successfully describe phenomena of the micro- and the
macro-cosmos respectively. So far they have resisted any attempt of complete
unification and quantum gravity \cite{kiefer_quantum_2007} is generally 
considered the Holy Grail of physics. Experimental tests of gravity 
\cite{will_confrontation_2006} with matter waves \cite{berman_atom_1997} 
started as early as 1975 with neutrons 
\cite{colella_observation_1975,rauch_neutron_2000}. 
Today, atom interferometers (AI) \cite{cronin_optics_2009} offer
new opportunities to probe the interface of these fundamentally disparate
descriptions of nature. The coherent evolution of quantum objects delocalized
in space-time \cite{lamine_ultimate_2006}, the verification of the Einstein 
principle of equivalence with quantum objects \cite{dimopoulos_testing_2007} 
and the detection of gravitational waves \cite{dimopoulos_atomic_2008}
constitute only three of many timely quests motivating experiments with AI in
extended free fall. The overarching aim is to enhance the sensitivity of these
devices, which increases linearly with the momentum difference between the two
matter waves \cite{chiow_102k_2011} emerging from a beam splitter and 
quadratically with the time of free fall as experienced in fountains 
\cite{dimopoulos_testing_2007,muller_atom-interferometry_2008}, drop towers 
\cite{van_zoest_bose-einstein_2010}, parabolic flights 
\cite{geiger_detecting_2011} and space 
\cite{international_school_of_physics_enrico_fermi_atom_2009}.  
These scaling laws imply constraints with respect to the atomic source. 
Thanks to their slow spreading and their excellent mode properties, 
Bose-Einstein condensates (BECs) \cite{cornell_nobel_2002,ketterle_nobel_2002} 
represent a promising source \cite{lamine_ultimate_2006} for high-resolution 
interferometers 
\cite{torii_mach-zehnder_2000,simsarian_imaging_2000,debs_cold-atom_2011}. 

In this letter, we report on the demonstration of a BEC-interferometer in
microgravity. Taking advantage of the extended free fall provided at the
drop-tower of the Center of Applied Space Technology and Microgravity (ZARM) in
Bremen, we have been able to coherently split a BEC consisting of about $10^4$
$^{87}$Rb atoms and to separate the emerging wave packets over macroscopic 
scales in time and space. The interferometer extends over more than half a 
second and covers distances of millimeters, exceeding the width of the 
condensate by an order of magnitude. By applying delta-kick cooling (DKC) 
\cite{chu_proposal_1986,ammann_delta_1997,morinaga_manipulation_1999} we have 
been able to reduce the expansion and to enhance the signal at longer 
interferometry times.

\begin{figure*}
\centerline{\includegraphics[width=13cm, natwidth=162bp,natheight=227bp]
{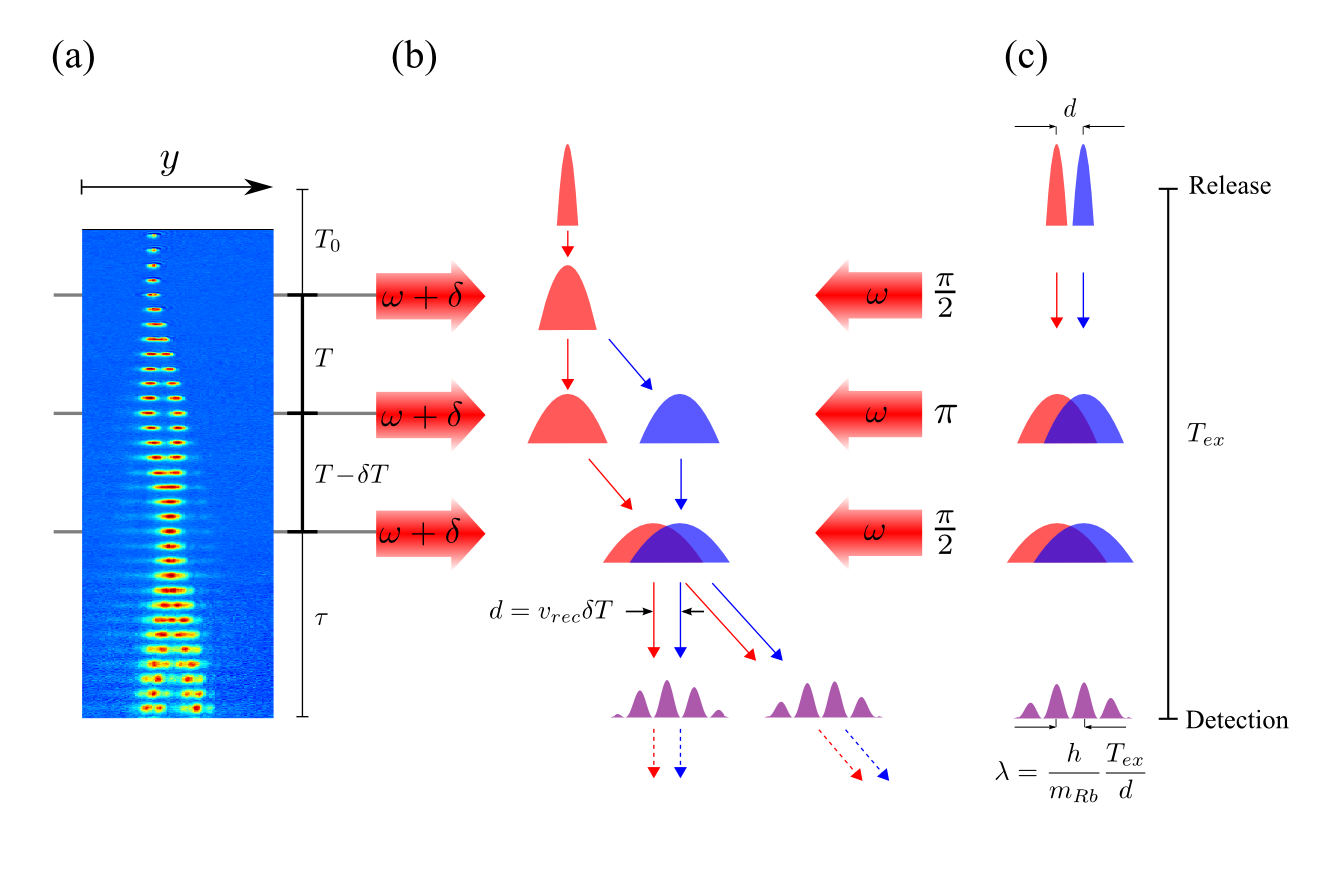}}
\caption{
Temporal AMZI for a BEC based on Bragg scattering at a light grating: experimental 
images on ground ({a}), schematic sequence ({b}) and analogy to 
the Young double-slit experiment ({c}). The evolution of the BEC and 
the AMZI is visualized by a series of absorption images ({a}) of the 
atomic densities separated by 1~ms. The incomplete transfer is a consequence 
of the larger mean field energy of the BEC necessary for the ground experiment. 
The interferometer starts at the time $T_0$ after the release of the BEC, when 
a $\pi/2$-pulse ({b}) made out of two counter-propagating light beams 
of frequency $\omega$ and $\omega+\delta$ creates a coherent superposition 
of two wave packets which drift apart with the two-photon recoil velocity 
$v_{rec} = $ 11.8~mm/s. After $T$ they are re-directed by a $\pi$-pulse and 
partially recombined after $T - \delta T$ by a second $\pi/2$-pulse. 
A nonzero value of $\delta T$  leads to a spatial interference pattern, 
which we record after $\tau = $ 53~ms in free fall. Similar to the far-field 
pattern observed in the Young double-slit experiment ({c}), the 
fringe spacing $\lambda$ scales linearly with the time of expansion 
$T_{ex} = T_0 + 2T - \delta T + \tau$  and is inversely proportional to the 
separation $d = v_{rec}\, \delta T$ of the wave packets. The scaling factor 
is the ratio of Planck's constant $h$ and the mass $m_{Rb}$ of the 
Rubidium atoms.}
\label{fig1}
\end{figure*}

 Our experiments are performed with an asymmetric Mach-Zehnder interferometer
(AMZI) \cite{torii_mach-zehnder_2000, rauch_measurement_1996} shown in 
Fig.\,1. Here we display the temporal evolution of the
atomic density distribution of the BEC-interferometer for an experiment on
ground [Fig.\,1(a)] and the corresponding experimental sequence for forming the
interferometer [Fig.\,1(b)]. A macroscopic wave packet is coherently split,
redirected and brought to a partial overlap by successive Bragg scattering at
moving light-crystals 
\cite{kozuma_coherent_1999,wang_atom_2005,nakagawa_atom_2009}.  
They are generated by pulses of two counter-propagating laser beams separated 
by the two-photon recoil energy of 15~kHz 
and detuned from the F=2 $\rightarrow$ F=3 transition of the $D_2$ line of 
$^{87}$Rb by 800~MHz to reduce spontaneous scattering. There exists a close 
analogy to the Young double-slit experiment [Fig.\,1(c)], where one pair of 
overlapping BECs plays the role of a pair of coherent light waves emanating 
from two slits separated by a distance $d$. Similar to the resulting 
interference pattern in the far-field of the double-slit, the fringe spacing 
in our expanding cloud, being the distance between two local maxima of the 
density, increases with the total expansion time $T_{ex}$ of the BEC and is 
inversely proportional to the displacement $d$ of the two clouds.

Figure~2 illustrates our experiments in microgravity performed at the drop
tower. We show the complete temporal sequence [Fig.\,2(a)] which differs 
from the previous experiments with the apparatus 
\cite{van_zoest_bose-einstein_2010} in three important features: (i)
We employ DKC to reduce the expansion during the free fall by briefly (2~ms)
switching on the trap with frequencies of (10, 22, 27) Hz generated by the atom
chip 30~ms after the release. (ii) In order to eliminate detrimental effects of
residual magnetic fields we transfer the BEC into the non-magnetic state 
$\left|F=2,m_F=0\right>$ by coupling the Zeeman levels with a chirped 
radio-frequency pulse (adiabatic rapid passage). (iii) At the time $T_0$  
after the release of the BEC, we implement the sequence of the AMZI outlined 
in Fig.\,1 and detect the interference pattern at $T_{ex}$ after the release 
using absorption imaging with a single laser pulse as illustrated in 
Fig.\,2(b). In Fig.\,2(c) we show typical images of the interfering BECs and 
the corresponding column density profiles for two different values of $T_{ex}$.

\begin{figure}
\centerline{\includegraphics[width=8cm, natwidth=162bp,natheight=227bp]
{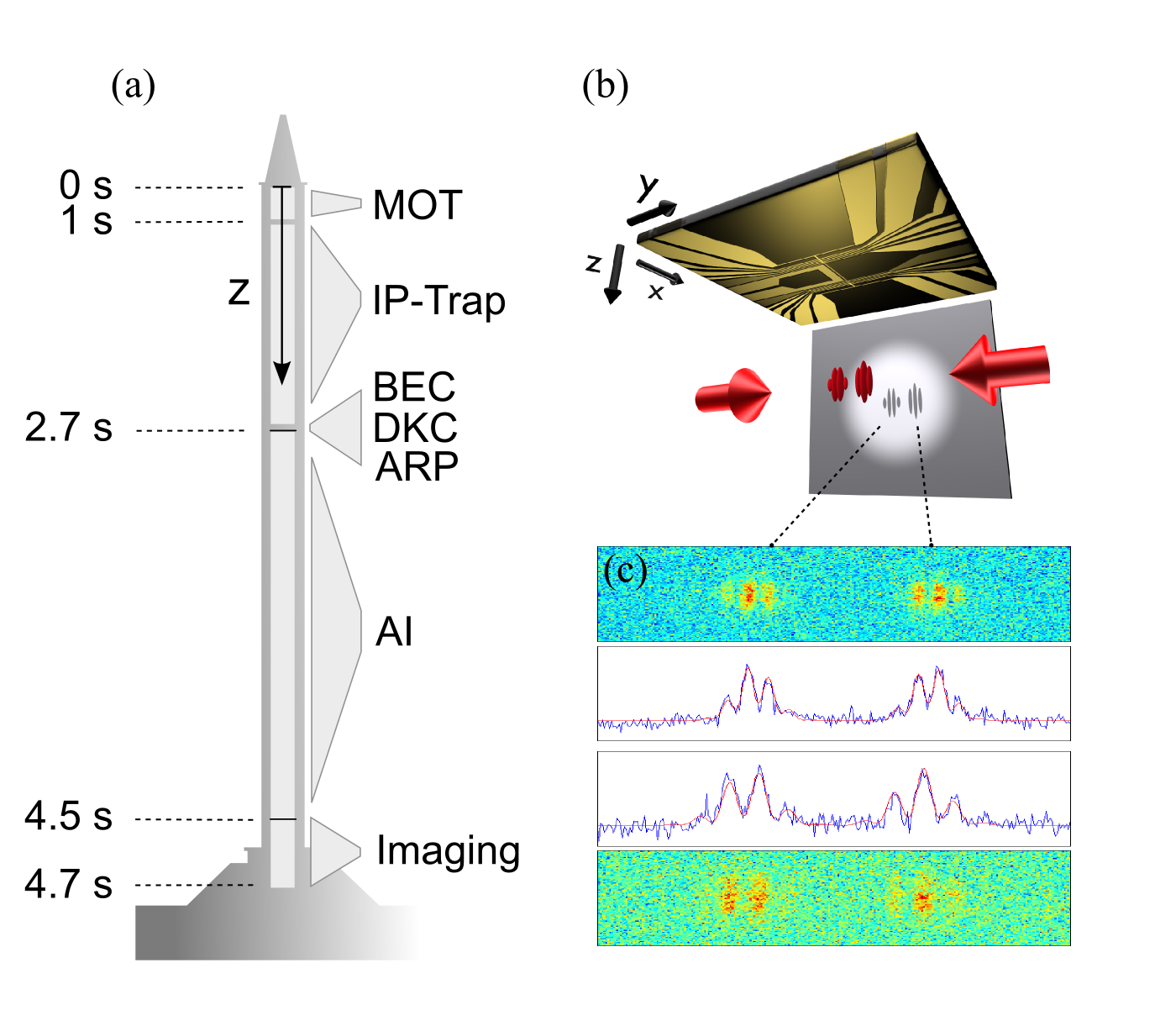}}
\caption{Mach-Zehnder interferometry of a BEC 
in microgravity as realized in the ZARM drop-tower in Bremen ({a}) 
where absorption imaging ({b}) brings out the 
interference fringes ({c}). The preparatory experimental 
sequence ({a}) includes capturing cold atoms in a magneto-optical 
trap (MOT), loading a Ioffe-Pritchard trap, creating a BEC and 
applying the DKC followed by the 
adiabatic rapid passage (ARP). The remaining time before the capture of the 
capsule at the bottom of the tower is used for atom interferometry (AI) and 
imaging of the atoms. The AMZI below the atom chip (top plane of {b}) 
is formed by scattering the BEC off moving Bragg gratings generated by two 
counter-propagating laser beams (red arrows directed along the y-axis), 
resulting in two pairs of interfering BECs. A resonant laser beam propagating 
along the x-axis projects the shadow of the BEC onto a CCD camera. 
Typical interference patterns and the corresponding column 
densities ({c}) are shown for $T_{ex}$ of 180~ms and 260~ms with 
corresponding fringe spacing of 75~\textmu m and 107~\textmu m.}
\label{fig2}
\end{figure}

\pagebreak
Figure~3 summarizes the central results of our paper on probing the coherent
evolution of a BEC with an interferometer in extended free fall. It shows the
spatial period of the observed fringe pattern [Fig.\,3(a)] as a function of the
expansion time $T_{ex}$, and the contrast [Fig.\,3(b)] observed at the 
exit ports of the AI for increasing values $2T - \delta T$ of the time the 
BEC spends in the interferometer. Moreover, in Fig.\,3(a) we confront the 
experimental results (blue circles, red squares, black triangles) with the 
corresponding theoretical predictions (solid blue and red lines). 
The solid blue line originates from a model based on the scaling approach 
\cite{SM} and describes the interference pattern of two condensates initially 
separated by a distance $d$, which start to expand
and eventually overlap. Their initial shape is derived from a detailed
numerical model of our magnetic chip trap. For large time scales the observed
fringe spacing (blue dots) shows a linear increase with $T_{ex}$ in full 
accordance with our model and with the linear far-field prediction 
(dash-dotted blue curve) of the double-slit. We emphasize, that our 
microgravity experiments operate deep in the linear regime and the non-linear 
behavior typical for the near-field combined with the non-linear evolution of 
the BEC occurs only at very short times ($<\,$30~ms). The linear scaling of 
the fringe pattern confirms the unperturbed evolution of the BEC during 
extended free fall.

\begin{figure*}
\centerline{\includegraphics[width=13cm, natwidth=162bp,natheight=227bp]
{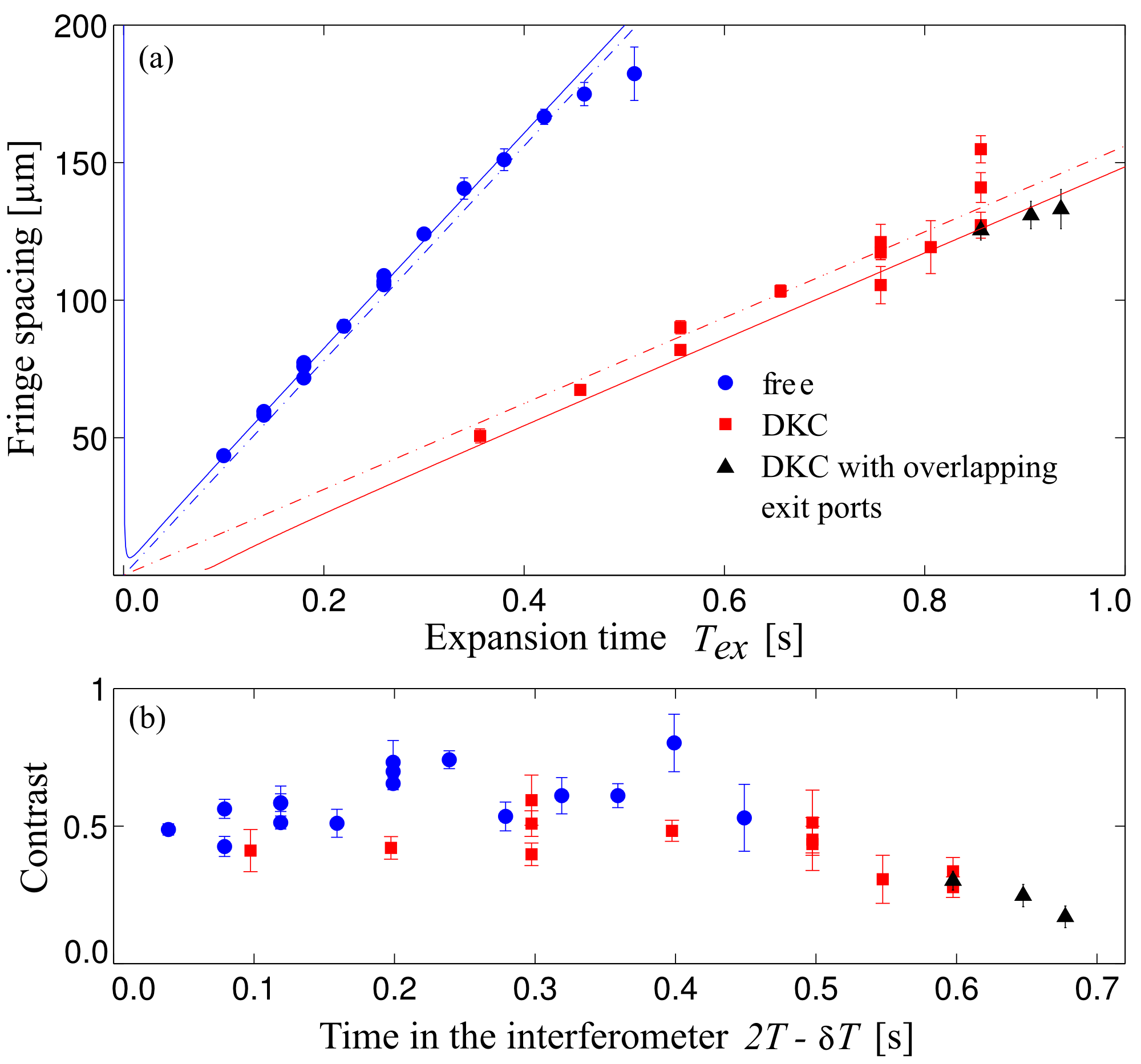}}
\caption{Fringe spacing ({a}) of two interfering BECs observed at each 
exit port of an AMZI with (red squares, black triangles, solid red line) and 
without (blue dots, solid blue line) delta-kick cooling (DKC) as a function of 
the expansion time $T_{ex}$ , and contrast ({b}) in its dependence on 
the time $2T - \delta T$  spent in the interferometer. Our data 
(blue dots, red boxes, black triangles) agrees well with our models of free 
expansion (solid blue line) and DKC (solid red line). In the case of DKC, the 
temporal asymmetry $\delta T$ was increased from 1~ms to 2.5~ms in order to 
preserve the number of visible fringes across the smaller BEC by rescaling the 
fringes, which decreased the slope in ({a}). With and without DKC, 
the observed linear dependence has the slope of the approximation of the 
double-slit for point particles (dash-dotted lines). An offset arises from the 
initial size and the nonlinear expansion of the BEC due to mean-field or from 
the focusing in DKC which shifts the apparent location of the double-slit to 
earlier or later times, respectively. The high contrast of the interference 
fringes ({b}), obtained by absorption imaging, typically exceeds 40\% 
but fades away for increasing $T$. Error bars depict 1-$\sigma$-confidence 
bounds of the fitted parameters.  }
\label{fig3}
\end{figure*}

The expansion rate of the BEC due to the mean-field energy is a limiting factor
for extending the interferometer to even longer time scales. It can be reduced
by DKC, which acts on the BEC like a three-dimensional lens. Indeed, in our
experiments DKC realized with the atom chip eliminates a substantial part of
the kinetic energy of the BEC, giving rise to an effective temperature of about
1nK. The method allows us to extend the observation of the free evolution of
the BEC and was tested with our AMZI. The experimental observations (red
squares) of the fringe spacing agree well with the theoretical predictions
(solid red line) for a double-slit experiment with delta-kick cooled atoms. In
order to reach even longer times (black triangles), we have adjusted the
detection time $\tau$ such that the patterns of the two exit ports overlap, 
thus increasing the absorption signal. 

As shown in Fig.\,3(b), we observe a contrast of more than 40\%, even at 
times $2T- \delta T$ as large as half a second. However, then the contrast 
decreases with the time over which the wave packets are separated, and 
generally with the expansion time of the BEC. The observed reduction is 
non-exponential in time and uniform over the cloud. In this respect the 
asymmetric interferometer puts more severe constraints on the set up than 
the symmetric one. However, it allows us to analyze various effects perturbing 
the interferometer. A preliminary analysis shows that the reduction may be due 
to an imperfect alignment of the beam splitters, inhomogeneous wavefronts or 
disturbances resulting from a slight capsule rotation. A more detailed 
discussion is subject to future investigations.

In conclusion, we have followed the evolution of the temporal coherence of an
asymmetric Mach-Zehnder interferometer with a BEC in microgravity, which is
analogous to Young's double-slit experiment with a gigantic matter wave packet
being in free fall for nearly a second. All preparatory steps necessary for
high-precision interferometry are implemented with the help of a robust
chip-based BEC source. Our device represents a unique testbed for exploring
atom interferometry with novel states of matter in extended free fall. In
particular, it allows us to test tools of atom optics, such as the precise mode
control of BEC with DKC at energy scales approaching pK temperatures. These
concepts are essential for high-resolution measurements both in fountains and
in microgravity. Moreover, the free fall of our apparatus results in a rather
compact interferometer. Indeed, a fountain-like interferometer would need more
than hundred meters to achieve the time of free fall provided by the catapult
at the ZARM. Last but not least, our  experiment  paves  the  way  for  quantum
tests  of  the  weak  equivalence  principle on sounding  rockets,  the
International  Space  Station and satellite missions \cite{STE-QUEST} 
such as STE-QUEST.


%





\begin{acknowledgments}
This project is supported by the German Space Agency (DLR) with funds provided 
by the Federal Ministry of Economics and Technology (BMWi) due to an enactment 
of the German Bundestag under grant numbers DLR 50 1131-1137 (project 
QUANTUS-III). The authors also thank the German Research Foundation (DFG) for 
funding the Cluster of Excellence QUEST Centre for Quantum Engineering and 
Space-Time Research.
\end{acknowledgments}

\nocite{camparo_dressed_1984,castin_bose-einstein_1996,storey_closed_2000,%
nandi_dropping_2007}
\bibliography{muentarxiv}

\pagebreak

\begin{appendix}

\pagenumbering{arabic}\renewcommand{\thepage}{A\arabic{page}}
\renewcommand{\thefigure}{A\arabic{figure}}

\section{Appendix}

\begin{center}
  \large{Supplementary Material for}\\
  \vspace{0.7cm}
  \Large{\textbf{Interferometry with Bose-Einstein Condensates\\in Microgravity}}
\end{center}

\section*{Materials and methods}
\label{sec1}

In this supplementary material, more details about key experimental
steps and the theoretical description are given. In the experimental part we 
provide details on the drop tower, the generation of the light grating 
for splitting and recombining the atomic ensembles, the state preparation \& 
detection and the quantitative analysis of the absorption images of fringe patterns. 

The theoretical description of the experimental results starts with a section
briefly reviewing the scaling approach as a tool for describing the expansion
dynamics of a Bose-Einstein condensate (BEC) for a time-dependent trapping
potential. This is then applied to the asymmetric Mach-Zehnder interferometer
(AMZI) of our experiment and all the relevant features of the interference
fringes at the exit ports are discussed. The complementarity of the interference
patterns of the two exit ports as well as the case of the detection of overlapping ports
are discussed in two separate sections.
\subsection*{Drop tower}
\label{droptower}

Our free-fall experiments are carried out in the drop tower of the
Center of Applied Space Technology and Microgravity in Bremen, Germany.
Inside the 146~m tall tower, a 122~m high steel tube 3.5~m in diameter
can be evacuated to around 10~Pa within 90 minutes. After a free fall of
4.7~s with residual accelerations in the parts per million range of the local gravity $g$, the sealed
experimental capsule is caught in a 8~m high pool of polysterene balls.
During the impact, the experiment has to sustain a deceleration of up to
50~$g$. A catapult mechanism housed beneath the tower allows us to launch
a capsule upwards and thus double the free-fall time, which we plan to use 
in future experiments.

\subsection*{Light-induced Bragg scattering}
\label{bragg}
Two light fields employed for Bragg scattering are generated by a single distributed
feedback laser diode, which is stabilized with a detuning of 800~MHz 
to the $F=2 \rightarrow  F^\prime=3$ transition in the D$_2$-line of $^{87}$Rb. 
The two beams pass an acousto-optic modulator 
(AOM) to generate the required relative offset frequency of 15~kHz.
This technique provides us with precise control of the timing and the shape 
of the pulses. The light is then coupled into two polarization-maintaining fibers 
and guided to the vacuum chamber, where the two counterpropagating collimated beams of
parallel linear polarization and diameter of 0.65~cm (FWHM) are formed.
They are aligned parallel to the chip surface and 
perpendicular to the detection axis so that the interference fringes can be 
detected by means of absorption imaging.
In the data presented, we use box-shaped pulses and Gaussian-shaped pulses.
The experiments showed that the pulse shape does not influence the contrast 
nor the signal of the interference pattern in the presented data. For this reason, we did not 
distinguish between the pulse shapes in Fig.~3.
\subsection*{State preparation }
\label{stateprep}
For the preparation of the BEC in the non-magnetic state $\left|F=2,m_F=0\right>$ 
we an adiabatic rapid passage$^\emph{31}$ (ARP).
For the
state transfer we apply a homogeneous static magnetic field of 11~G in
$x$-direction and combine this with a 3.8~ms long RF-sweep from 7.714~MHz to
7.754~MHz. After the interferometer sequence a quadrupole field in Stern-Gerlach
configuration is applied to split the atoms according to their hyperfine-state
and to ensure that only atoms in the non-magnetic state contribute to the
interference pattern. In this way we are also able to verify transfer
efficiencies to the non-magnetic state of about 90\%. 

\subsection*{Detection and analysis of absorption images}
\label{analysis}

The atomic clouds are probed with a laser beam on the $F = 2 \rightarrow
F^\prime = 3$ transition where the intensity of the detection beam is typically 
about 5\% of the saturation intensity of the transition. 
Two images are taken in each experiment with a charge-coupled-device (CCD) camera (Hamamatsu
C8484-15G):
First the shadow of the cloud  in the detection beam is recorded providing us 
with the intensity $I_{atoms}$. This beam heats the atoms and expels them from 
the imaging region. Then the detection beam is imaged in the absence of any 
atoms yielding $I_{beam}$. Both intensities $I_{atoms}$ and $I_{beam}$ are 
corrected for the camera dark image $I_{dark}$ and we obtain the optical density
\begin{equation}
  D \equiv \ln \left( \frac{ I_{beam} - I_{dark}}{I_{atoms} - I_{dark}} \right) 
  \label{eq:opticaldensity},
\end{equation}
and the atomic column density
\begin{equation}
  n \equiv \frac{D}{\sigma}
  \label{eq:columndensity}
\end{equation}
with the resonant cross section $\sigma$.

To extract the contrast and the fringe spacing from the images of the interferometer
output ports, we sum the column densities along $z$ and fit the function
\begin{align}\label{eq:1dfringefit}
  n_{1D} \equiv ~&n_{max,1D} \left[ 1+C \sin\left( \frac{2\pi}{\lambda} (y-y_1)+\varphi \right) \right]   \exp \left( -\frac{(y-y_1)^2}{2 \sigma_y^2} \right)  \notag\\
  &+ n_{max,1D} \left[ 1+C \sin\left( \frac{2\pi}{\lambda} (y-y_2)+\varphi +\pi \right) \right]   \exp \left( -\frac{(y-y_2)^2}{2 \sigma_y^2} \right)  +n_{0,1D}
\end{align}
to the 1D profiles with a nonlinear least squares method. Here, $y_1$ and $y_2$ 
are the centers of the two output ports, $\sigma_y$ the width of their Gaussian envelopes, 
$\lambda$ the fringe spacing and $C$ the contrast shown in Fig.3(a) and Fig.3(b) respectively, 
while $n_{0,1D}$ accounts for an offset due to intensity fluctuations between the images. 
The phase $\varphi$ would be read out for an interferometric measurement of e.g. inertial forces.
For long observation times a tilt of the interference fringes with respect to the $z$-axis is observed, which is compensated 
by aligning the image before we calculate the column density along $z$. Error bars in Fig.3 are standard deviations calculated 
via the fitted model and multiple noise realizations.

\begin{figure}[!h]
  \begin{center}
    \includegraphics[width=.8\textwidth]{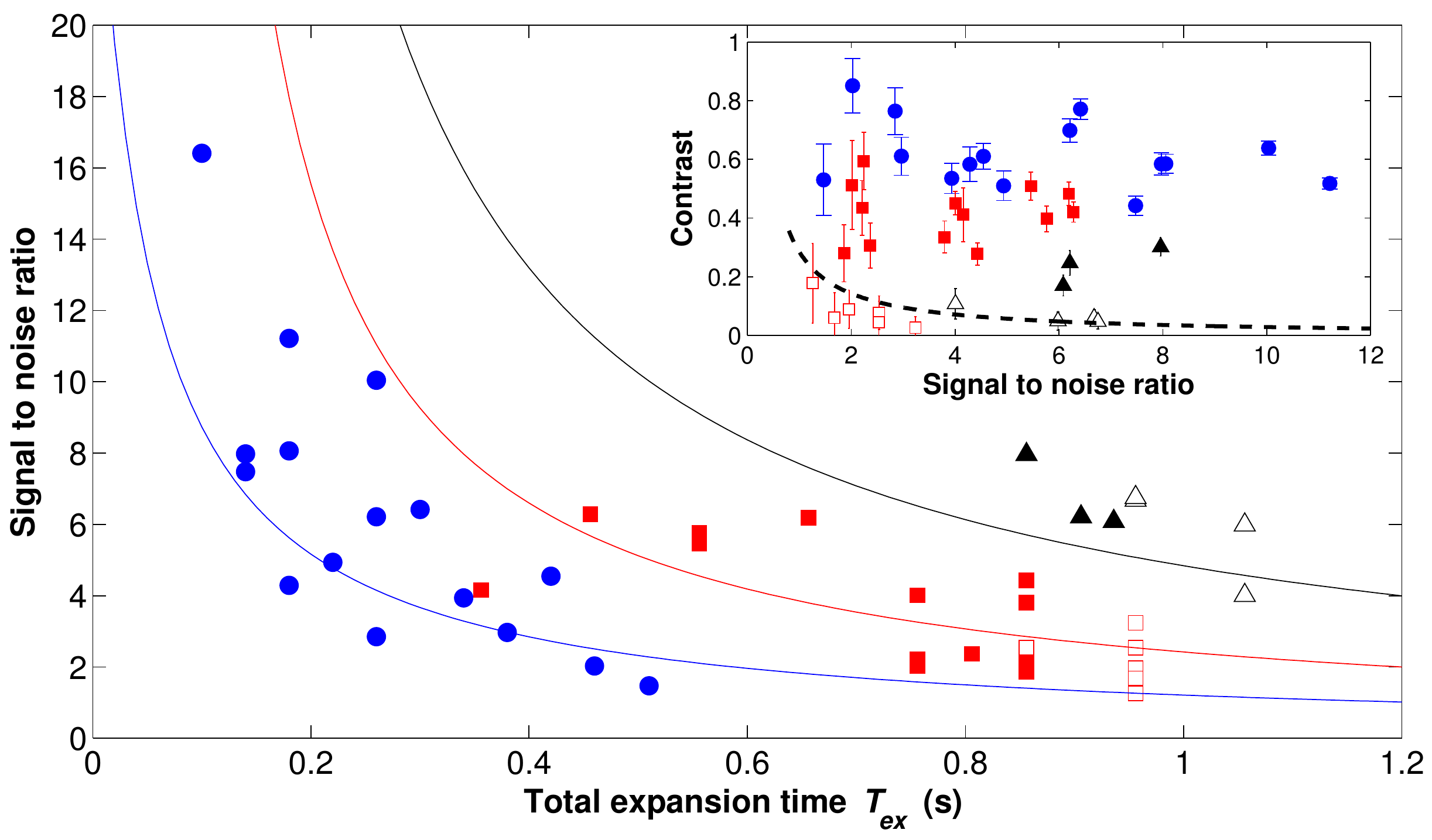}
  \end{center}
  \caption{Signal to noise ratio (SNR) of the integrated atomic density forming 
  the interference patterns as a function of the total expansion time $T_ {ex}$ 
  with (red squares) and without (blue dots) delta kick cooling (DKC) as well as for 
  overlapping interferometer ports (black triangles). With DKC the SNR decreases 
  with a slower rate than without and is further increased when both interferometer ports 
  are detected before they are fully separated spatially. The scatter is caused by 
  atom number fluctuations and by the detection intensity varying between individual 
  drops due to the atoms being located in another position within the Gaussian 
  profile of the laser beam. To facilitate the comparison of the  three cases, the 
  curves (blue, red, black) show the decay according to  Eq.\eqref{eq:SNR} and \eqref{eq:sigdecay} using 
  the measured expansion rates, mean atom number and noise. Towards longer expansion 
  times (open black triangles and red squares), the period of the fringe pattern 
  could not be obtained by fitting the data with the period as a free parameter, 
  so these data points were discarded in Fig.~3.  
  The inset presents the observed contrast in relation to the SNR for all datapoints. 
  The contrast was obtained by fitting the function $n_{1D}$ given by. 
  Eq.~\eqref{eq:1dfringefit} to the experimental data (blue circles, red squares 
  and black triangles) and to synthetic data (black dashed line) generated 
  from Eq.~\eqref{eq:1dfringefit} for a vanishing contrast ($C=0$) and noise in 
  compliance with the observed SNR. For the synthetic data as well as for the 
  discarded experimental data points the contrast was derived from fits with the 
  fringe period fixed to the theoretical value. The agreement of both confirms the 
  loss of contrast for long interferometer times $2T - \delta T$. For smaller SNRs, 
  the minimum detectable contrast increases, as indicated by the dashed line.
}

  \label{fig:SNR}
\end{figure}

Figure \ref{fig:SNR} displays the decaying signal to noise ratio 

\begin{equation}\label{eq:SNR}
  S\!N\!R  \equiv \frac{n_{max,1D}}{\nu_{rms}}
\end{equation}

of the integrated atomic density with and without delta kick cooling (DKC) as a function of time, with $\nu_{rms}$ being 
the RMS of the fit residuals. Due to the spreading of the atomic wave packet, in the linear regime the 1D atomic density decreases according to

\begin{equation}\label{eq:sigdecay}
 n_{max,1D} = \frac{N}{\sqrt{8\pi} \sigma_y} \approx \frac{N}{\sqrt{8\pi} \left( \alpha T_{ex} +\beta \right)}
\end{equation}

with the total atom number $N$, expansion rate $\alpha$ and an offset $\beta$.
Thus, in the case of DKC the SNR falls off with a slower rate and is doubled by overlapping the output ports, allowing us to detect the atoms after longer times $T_{ex}$ than in the free case.
 
For longer interferometer times $2T-\delta T$ the fit routine does not converge 
to the expected fringe period, which is kept as a free parameter in the data evaluation.
These data (open red squares and open black triangles in Fig.~\ref{fig:SNR} ) 
were not included in Fig.~3. Fitting the discarded data with the fringe period 
predicted by Eq. \eqref{eq:period} yields a finite contrast. These values
were confronted with the contrast obtained by applying the same fit routine to 
synthetic test signals, which were generated from Eq.~\ref{eq:1dfringefit} for a 
vanishing contrast ($C=0$) with multiple noise realizations in compliance with the observed SNR. 
As depicted in the inset of Fig.~\ref{fig:SNR}, the contrast resulting from the
analysis of the test data (dashed line) was indeed equal or higher than the one
obtained for the discarded experimental data, thus confirming the loss of contrast. 
The dashed curve indicates that the minimum detectable contrast is limited by the SNR.

\clearpage

\subsection*{Asymmetric Mach-Zehnder interferometer}

We now turn to the theoretical description of our asymmetric Mach-Zehnder interferometer (AMZI). Here we only outline the essential 
ingredients of our formalism and present the main results relevant for the experiments described in our article. A more detailed treatment will be subject of a future publication.

\subsubsection*{Building blocks}
We describe the BEC in our trap by a time- and position-dependent macroscopic wave function $\psi=\psi(\vec{r},t)$ determined 
by the non-linear Gross-Pitaevskii equation (GPE). In all experiments reported in our article the trap is well approximated by a 
parabolic potential $V$ which is also time-dependent since the BEC is released by turning off the trap. Another reason for 
considering a time-dependent $V$ emerges from the need to describe the technique of delta-kick cooling in our experiments.
In these cases, the GPE can be well approximated by an almost analytical solution $\psi = \E{\I \Phi} \psi^{\text{(TF)}}(\vec{r},t)$ 
consisting of the product of a phase factor and the Thomas-Fermi (TF) ground state wave function of the trap with a time-dependent 
rescaling of the spatial coordinates$^\emph{14, 32-34}$.
The so-defined time-dependent wave function $\psi^{\text{(TF)}}(\vec{r},t)$ is real and normalized at all times. 
Moreover, the phase $\Phi$ of $\psi$ does not only depend on the space-time variables $\vec{r}$ and $t$ but also 
on the center-of-mass (COM) motion of the BEC expressed by the time-dependent coordinate $\R=\R(t)$ and momentum $\sP=\sP(t)$. 
In contrast, the TF-wave function $\psi^{\text{(TF)}}$ is only a function of the difference $\vec{r}-\R(t)$ and $t$.

The interaction of the BEC with a Bragg pulse at a time $t_\mathrm{p}$ can change the instantaneous 
momentum $\sP(t_\mathrm{p})$ of the COM motion by $\pm \hbar \vec{k}$, where $\vec{k}$ is the effective 
wave vector of the Bragg pulse, but leaves the position $\R(t_\mathrm{p})$ untouched. Moreover, the wave 
imprints a phase on the BEC given by the sum of $\vec{k} \cdot \R(t_\mathrm{p})$ and the phase $\phi (t_\mathrm{p})$ 
of the Bragg pulse. The wave function altered in this way is also multiplied by the imaginary factor $(-\I)$ resulting from the time-dependent Schr{\"o}dinger equation.

When the parameters of the Bragg pulse are chosen to act as a 50:50 beam splitter, the BEC partly experiences 
a recoil and partly continues undisturbed with equal amplitudes. As a result, the COM motion is in a superposition 
of the original momentum $\sP$ and $\sP+ \hbar \vec{k}$, or $\sP$ and $\sP - \hbar \vec{k}$ with equal 
weights $1/\sqrt{2}$. When the pulse acts as a mirror, it changes the momentum by $+\hbar \vec{k}$ or $-\hbar \vec{k}$ with unit probability.

\subsubsection*{Wave functions at the exit ports}
\begin{figure}[h]
\begin{center}
\includegraphics[scale=1]{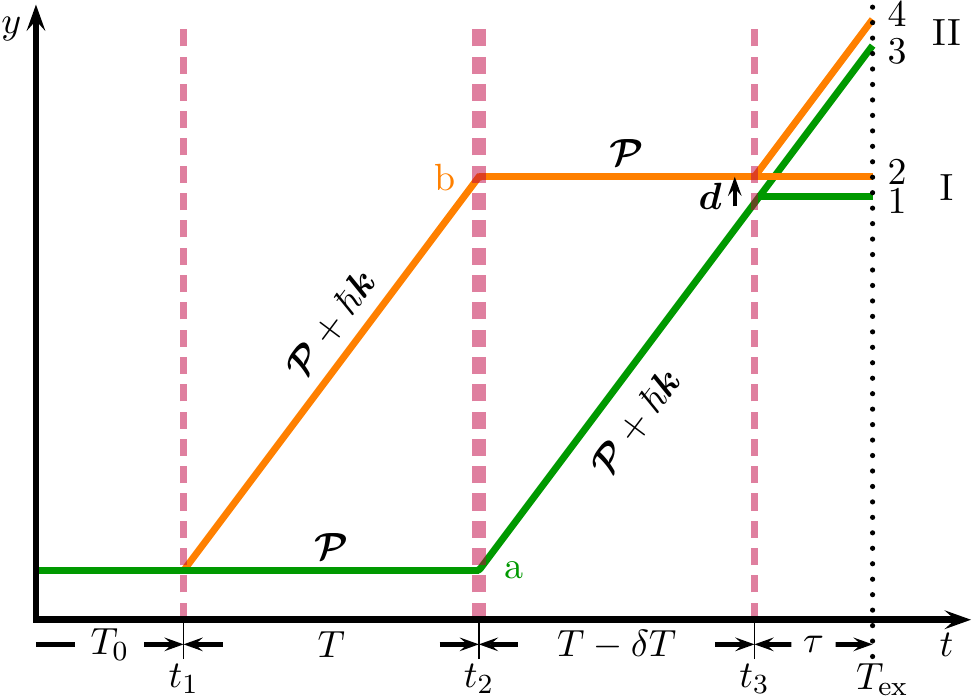}
\end{center}
\caption{
Diagram of the COM trajectories for our AMZI with labels for the two branches (a and b), the two exit ports (I and II) and the equal-momentum trajectories at each exit port ($\{1,2\}$ and $\{3,4\}$).
The horizontal axis corresponds to time and the vertical one to position along the $y$-direction.
The dashed lines depict the Bragg laser pulses whereas the dotted line represents the detection pulse for absorption imaging.}
\label{fig:AMZI}
\end{figure}

The description of the BEC with the help of the GPE and the effect of the Bragg pulses outlined above constitute the main building blocks of our formalism to analyze the AMZI shown in Fig.~\ref{fig:AMZI}. Our goal is to obtain the wave functions $\Psi_{\text{I}}$ and $\Psi_{\text{II}}$ of the BEC at the two exit ports I and II of the interferometer, which differ by the COM momentum of the BEC.

Indeed, in port I it is the initial momentum $\sP$ since both paths 1 and 2 have experienced two opposite momentum kicks. However, in port II it is $\sP + \hbar \vec{k}$ since both paths have received an odd number of kicks which is different in paths 3 and 4. In path 4 we have three exchanges of momenta, that is $\hbar \vec{k}-\hbar \vec{k}+\hbar \vec{k}=\hbar \vec{k}$, whereas in path 3 we have only a single one, that is $+\hbar \vec{k}$.

In order to find the wave functions $\Psi_{\text{I}}$ and $\Psi_{\text{II}}$ at the two exit ports, we now solve the GPE including the momentum kicks imparted by the Bragg pulses starting with the initial TF-wave function of the ground state of the trap. We switch off the trap at $t=0$ and evolve the BEC according to the GPE.
This dynamics gives rise to a phase factor $\Phi_0$ and a spreading of the TF-wave function. 
The Bragg pulse at $t_1\equiv T_0$ serving as a beam splitter either changes the COM motion due to the momentum transfer $\hbar \vec{k}$ and imprints phases on the BEC in addition to the $\Phi_0$ accumulated before the pulses, or leaves it untouched. After the pulse the two BECs evolve for the time $T$ according to the GPE in the absence of a potential, which provides us with additional dynamical phases.

The second Bragg pulse at $t_{2} \equiv T_0 + T$ acts as a mirror and imprints again a phase but also transfers momentum with unit probability. Hence, no splitting of the trajectories occurs. The ensuing dynamics of the BECs according to the GPE adds new dynamical phases, and the two BECs on the paths a and b approach each other till at $t_3\equiv T_0+2T-\delta T$ the third Bragg pulse implementing another beam splitter mixes them. Again the phases of the pulse and the instantaneous positions of the BEC are mapped onto the wave functions.

At this moment it becomes important that our interferometer is asymmetric since the time $T$ between the first and the second Bragg pulse is different from the one between the second and the third by an amount $\delta T$, as indicated in Fig.~\ref{fig:AMZI}. As a consequence, at each exit port the interfering BECs are slightly shifted with respect to each other by an amount $\vec{d}\equiv (\hbar \vec{k}/m) \delta T$ where $m$ is the mass of the atom. 

When this displacement $\vec{d}$ is small compared to the extension of the BEC, we can neglect it in the arguments of $\psi^{\text{(TF)}}$, which acts as envelope of the wave functions contributing to each exit port. However, it is of crucial importance for the phases accumulated by the BEC on the two paths through the interferometer, since they depend in a sensitive way on the COM motion.

An execution of this procedure yields for $t_3<t$ the expressions
\begin{eqnarray}
\Psi_{\text{I}} \cong \frac{1}{2} \E{\I \alpha_{\text{I}}} \psi^{\text{(TF)}}\left(\vec{r}-\R_1(t) ; t\right) \left\lbrace 1+\ehoch{\I \left(\alpha +\frac{m\vec{d}}{\hbar t} \left(\vec{r}-\R_1(t)\right)\right)} \right\rbrace \label{eq:Psi_I}
\end{eqnarray} 
and
\begin{eqnarray}
\Psi_{\text{II}} \cong \frac{1}{2} \E{\I \alpha_{\text{II}}} \psi^{\text{(TF)}}\left(\vec{r}-\R_3(t) ; t\right) \left\lbrace 1-\ehoch{\I \left(\alpha +\frac{m\vec{d}}{\hbar t} \left(\vec{r}-\R_3(t)\right)\right)} \right\rbrace \label{eq:Psi_II}
\end{eqnarray}
for the wave functions $\Psi_{\text{I}}$ and $\Psi_{\text{II}}$ in the two exit ports. Here $\alpha_{\text{I}}$, $\alpha_{\text{II}}$ and $\alpha$ contain the phases imprinted by the Bragg pulses as well as the dynamical phases accumulated by the BEC along the two paths and determined by the GPE.

The two wave functions are centered around the positions $\R_1(t)$ and $\R_3(t)$ of the COM trajectory at $t$ and the spatially dependent phase, leads to oscillations with wavelength
\begin{eqnarray}
\lambda \equiv \frac{2 \pi \hbar t}{m|\vec{d}| } ~.\label{eq:period}
\end{eqnarray}

\subsubsection*{BEC densities: Near-field versus far-field}

From Fig.~\ref{fig:AMZI} we note that at the final beam splitter, that is at $t_3$, we have the equality $\R_1(t_3)=\R_3(t_3)$, and due to the momentum difference of $\hbar \vec{k}$ between $\R_3$ and $\R_1$ we find the relation
\b
\R_3(t_3+\tau)= \R_1(t_3)+\frac{\hbar \vec{k}}{m} \tau
\e
for later times.

When $(\hbar |\vec{k}|/m)\tau$ is larger than the extension of the BEC given by $\psi^{\text{(TF)}}$, we can separate the two exits and the BEC densities $W_{\text{I}}$ and $W_{\text{II}}$ following from \eqref{eq:Psi_I} and \eqref{eq:Psi_II} read 
\b
W_{\text{I}}=  \left( \psi^{\text{(TF)}}\left(\vec{r}-\R_1(t);t\right) \right)^2 \cos^2\left[\frac{\alpha}{2}  +\frac{m\vec{d}}{2\hbar t}\cdot \left(\vec{r}-\R_1(t)\right) \right]
\label{eq:portI}
\e
and
\b
W_{\text{II}}= \left( \psi^{\text{(TF)}}\left(\vec{r}-\R_3(t);t\right) \right)^2 \sin^2\left[\frac{\alpha}{2}  +\frac{m\vec{d}}{2\hbar t} \left(\vec{r}-\R_3(t)\right) \right] ~.
\label{eq:portII}
\e

Three features stand out most clearly in these expressions valid in the far-field limit:
($i$) The envelopes of $W_{\text{I}}$ and $W_{\text{II}}$ are given by the time-dependent TF-profiles $\left( \psi^{\text{(TF)}}\right)^2$ centered at the COM coordinates $\R_1(t)$ and $\R_3(t)$ of the BEC at $t$.
($ii$) The arguments of the trigonometric functions in $W_{\text{I}}$ and $W_{\text{II}}$, which are also measured with respect to $\R_1(t)$ and $\R_3(t)$, have identical wavelengths given by \eqref{eq:period} and depend on the total time from the release of the BEC from the trap till the measurement, denoted by $T_\mathrm{ex}$ in the paper, and 
($iii$) the two distributions are complementary, that is $W_{\text{I}}$ follows a cosine- and $W_{\text{II}}$ a sine-function.

It is also interesting to note that we would have obtained the same expression for $W_{\text{I}}$ if we had considered the free propagation of a superposition of two identical wave packets initially separated by $\vec{d}$ and with momentum $\sP$. Indeed, the fringe spacing $\lambda$ in both situations is identical. In the corresponding analogy giving rise to $W_{\text{II}}$ the wave packets have the momentum $\sP+\hbar \vec{k}$. It is in this sense that the AMZI is analogous to Young's double-slit configuration.

In the opposite limit, when $(\hbar |\vec{k}|/m)\tau$ is smaller than the extension of the BEC, the two wave functions $\Psi_{\text{I}}$ and $\Psi_{\text{II}}$ create an interference term $I\equiv \Psi_{\text{I}}^* \Psi_{\text{II}}+\, \text{c.c.}$ in the probability
\b
W= \left| \Psi_{\text{I}}+\Psi_{\text{II}}\right|^2= W_{\text{I}}+W_{\text{II}}+I~.
\e
From \eqref{eq:Psi_I} and \eqref{eq:Psi_II} we note that $I$ contains the phase difference $\alpha_{\text{I}}-\alpha_{\text{II}}$,
which apart from many other phases involves a term $\vec{k}\cdot \vec{r}$ due to the different COM momenta of the paths 3 and 1 after the last beam splitter and gives rise to oscillations of the term $I$ with the optical wavelength. Our detectors cannot resolve such small structures and average instead over them, so that the contributions of the interference term $I$ vanishes.

When we now arrange the detection time $\tau$  such that
\b
\frac{ \vec{d}\cdot \vec{k}\tau}{2t}= \frac{\pi}{2}~,
\e
we can make a  cosine- out of the sine-function in $W_{\text{II}}$, which amounts to a shift by $\lambda/2$ so that the maxima of $W_{\text{I}}$ and $W_{\text{II}}$ coincide.

Moreover, we recall that we are in the near-field regime where we cannot distinguish the two exit ports, that is we can replace the COM coordinate $\R_3$ by $\R_1$ in the prefactor $\psi^{\text{(TF)}}$ of $W_{\text{II}}$, which finally yields the expression
\b
W = 2 \left( \psi^{\text{(TF)}}\left(\vec{r}-\R_1(t);t \right)\right)^2 \cos^2\left[\frac{\alpha}{2} +\frac{m\vec{d} }{2\hbar t}\left(\vec{r}-\R_1(t)\right) \right]
\label{eq:overlap}
\e
for the BEC density.

Hence, by working in the near field region, we have obtained a factor of two in the intensity of $W$ which allowed us to obtain a higher signal and extend our measurements to longer times. In Fig.~3 such measurements are indicated by triangles.

\end{appendix}

\end{document}